
\magnification 1200
\baselineskip=16pt
\centerline{\bf Weakly decaying asymptotically flat static}
\centerline{\bf and stationary solutions to the Einstein equations}
\vskip 2cm
\centerline{Daniel Kennefick$^*$ and Niall \'O Murchadha}
\vskip 1cm
\centerline{\it Physics Department, University College, Cork, Ireland.}
\vskip 2cm
\centerline{\bf Abstract}
The assumption that a solution to the Einstein equations is static (or
stationary) very strongly constrains the asymptotic behaviour of the metric. It
is shown that one need only impose very weak differentiability and decay
conditions {\it a priori} on the metric for the field equations to force the
metric to be analytic near infinity and to have the standard Schwarzschildian
falloff.
\vfill\eject
\centerline{\bf I: Introduction}
It has long been known that the asymptotically flat static and stationary
solutions of the Einstein equations are analytic at spacelike infinity (Beig
and Simon 1980, 1981). This should come as no real surprise, because they are
the special solutions without gravitational radiation and so therefore the
far-field should be dominated by the Newtonian potential(s). Further, Beig and
Simon (1980, 1981) have shown that all static and stationary solutions are
Schwarzschildian at infinity.

In this article, we wish to improve on this result by weakening (in a
significant way) the assumptions made, yet recovering the Beig and Simon
results, both about analyticity and about falloff.

One assumption made by Beig and Simon was that the given (static or
stationary) metric went flat like 1/r (in some asymptotically cartesian
coordinate system). It has been shown that `reasonable' solutions to the
Einstein equations exist which fall off much slower at infinity
(Christodoulou and \'O Murchadha 1981). For example, the condition for finite
total energy is that the metric go flat faster than $r^{-{1 \over 2}}$
(Bartnik 1986, Chru\'sciel 1986a,b, \'O Murchadha 1986). With such a
fall-off, it can be shown that the energy is very well behaved, it is
positive and is a component of a well-defined timelike energy-momentum vector
which transforms correctly under Lorentz transformations. Christodoulou and
\'O Murchadha (1981) even discussed solutions which went flat essentially
arbitrarily slowly. Such solutions, despite the fact that their masses were
infinite, could reasonably be described as being asymptotically flat with
complete spacelike infinities.

One result of this paper is that all such slow decay at
infinity must be due to the existence of gravitational radiation near spacelike
infinity. In particular, we show that if we assume that the solution has slow
falloff near infinity and that it is simultaneously either static or stationary
then the field equations force the usual ${1 \over r}$ falloff on us and we
recover the standard Beig and Simon Schwarzschildian result.

In addition to allowing slow decay at spacelike infinity, we also assume that
the metric belongs to some Sobolev space and so need only be weakly
differentiable, rather than belonging to any classical space. Again, the field
equations force up the differentiability and we can finally conclude that the
metric must be analytic.

In this paper, we are careful to impose conditions only on the metric in the
physical spacetime. We then use the field equations to show that other
(unphysical) quantities are suitably smooth as we need them. We take a
straightforward 3 + 1 approach to the problem and write the static and
stationary field equations as conditions on the Cauchy data for the
gravitational field (Arnowitt, Deser and Misner 1962 (ADM)).
\vskip 1cm
\centerline{\bf II: Asymptotically flat static and stationary solutions}
\vskip 0.5cm
We will consistently adopt an initial value approach in this article. This
means that we assume we are given initial data for the gravitational field on a
single spacelike slice. The Einstein evolution equations now determine how this
data evolves. This evolution then can be tested to see whether the spacetime in
either static or stationary. We are only interested in working near infinity so
we will assume that the data is given on a region equivalent to $R^3$ with a
ball cut out.

Initial data for the vacuum gravitational field consists of a pair $(g_{ij},
\pi^{ij})$ (ADM), where $g_{ij}$ is an asymptotically flat Riemannian
three-metric on the given manifold and $\pi^{ij}$ is a symmetric three-tensor
density. These are not independent, they must satisfy the constraints
$$^{(3)}R = {1 \over g}[\pi^{ab}\pi_{ab} - {1 \over 2}(g_{ab}\pi^{ab})^2]
\eqno(2.1)$$
$$\nabla_i\pi^{ij} = 0 \eqno(2.2)    $$
where $^{(3)}R$ is the three scalar curvature and $\nabla$ is the covariant
derivative compatible with $g_{ij}$. The rest of the Einstein equations form
the evolution equations
$${\partial g_{ij} \over \partial t} = 2Ng^{-{1 \over 2}}[\pi_{ij} - {1 \over
2} (g_{ab}\pi^{ab})g_{ij}] + N_{i;j} + N_{j;i} \eqno(2.3)$$
$$\eqalignno{{\partial\pi^{ij} \over \partial t} &= -Ng^{{1 \over 2}}[R^{ij} -
{1 \over 2}g^{ij}R] + {1 \over 2}Ng^{-{1 \over 2}}g^{ij}[\pi_{ab}\pi^{ab}
- {1 \over 2}(g_{ab}\pi^{ab})^2] -2Ng^{-{1 \over 2}}[\pi^{im}\pi^j_m \cr
&\ \ \ - {1 \over 2}(g_{ab}\pi^{ab})\pi^{ij}] + g^{{1 \over 2}}(N^{;ij} -
g^{ij}\nabla^2N) + (\pi^{ij}N^m)_{;m} - N^i_{;m}\pi^{mj} -N^j_{;m}\pi^{im} &
(2.4)\cr} $$
The extra scalar and vector $(N, N^i)$ (the lapse and shift) represent the
arbitrariness in the choice of ``time'' off the hypersurface. That is
$$t^{\mu} = N n^{\mu} + N^{\mu}, \ \ \ n^{\mu}N_{\mu} = 0, \eqno(2.5)$$
where $t^{\mu}$ is the time translation vector and $n^{\mu}$ is the unit
(timelike) normal to the hypersurface; $N^{\mu}$ is really a three-vector, it
lies entirely in the given slice.

Now we wish to consider the case where the given data represents a slice
through a spacetime which has a timelike Killing vector $k^{\mu}$
(at least near infinity). This implies that for the given $(g, \pi)$ there
exists a particular choice of $(N, N^i)$ which, when substituted into (2.3) and
(2.4), gives
$${\partial g_{ij} \over \partial t} = {\partial\pi^{ij} \over \partial t} = 0.
\eqno(2.6)$$
This special choice of lapse and shift are those which correspond to using the
killing vector as ``time'' direction. In other words, we split up $k^{\mu}$,
as suggested by eq.(2.5), to generate the desired $(N, N^i)$,
$$k^{\mu} = N n^{\mu} + N^{\mu}. \eqno(2.7)$$
Let us further assume that the spacetime is static. This means that we can
assume that the killing vector is hypersurface orthogonal. Let us choose our
three-slice to be the hypersurface that is perpendicular to the killing
vector. This means that the killing vector is parallel to $n^{\mu}$ and the
shift $N^{\mu}$ will be zero. To repeat: for a static spacetime we must have
an initial data set ($g, \pi $) (satisfying the constraints) and a
choice of lapse $N$ (with $N^i = 0$) such that (2.6) holds. $N$ is just the
length of the killing vector.

In the light of this (2.3) reads
$$N[\pi_{ij} - {1 \over 2} (g_{ab}\pi^{ab})g_{ij}] = 0. \eqno(2.8) $$
Since we must have $N \not = 0$ (otherwise the Killing vector vanishes) we
immediately get
$$ \pi^{ij} \equiv 0. \eqno(2.9)$$
Obviously, (2.2) is trivially satisfied. The Hamiltonian constraint (Eq.
(2.1)) gives us
$$^{(3)}R \equiv 0. \eqno(2.10) $$
When (2.9) and (2.10) are substituted into (2.4) we get
$$0 = -N R^{ij} + (N^{;ij} - g^{ij} \nabla^2 N). \eqno(2.11)$$
The trace of (2.11) gives
$$0 = -N^{(3)}R - 2\nabla^2 N. \eqno(2.12)$$
But since $^{(3)}R = 0$ (Eq. (2.10)) we get
$$\nabla^2 N = 0. \eqno(2.13)$$
Finally, when (2.13) is substituted back into (2.11) we find that
$$N R_{ij} = N_{;ij}. \eqno(2.14)$$
Therefore a static solution to the Einstein field equations is equivalent to
having a Riemannian three-metric $g$ and a scalar function $N$ which satisfy
(2.14), together with either (2.13) or (2.10). We need only one or the other
because the trace of (2.14) gives
$$N^{(3)}R = \nabla^2N. \eqno(2.15)$$

Having obtained the static field equations in the form (2.10), (2.13), (2.14)
we wish to manipulate them further. We wish to use the lapse $N$ as a conformal
factor to simplify the equations. In particular we define a function $u$ by
$$e^u = N, \eqno(2.16) $$
and transform the metric by the rule
$$\tilde g_{ab} = N^2g_{ab} = e^{2u}g_{ab}. \eqno(2.17)$$
A standard calculation now shows
$$\tilde R_{jl} = R_{jl} - u_{;jl} -g_{jl}\nabla^2u + u_{;j}u_{;l} - (\nabla
u)^2g_{jl}. \eqno(2.18)$$
It is easy to show
$${N_{;ij} \over N} = u_{;ij} + u_{;i}u_{;j},$$
$$\nabla^2N = e^u[\nabla^2u + (\nabla u)^2],$$
$$\tilde \nabla^2u = e^{-2u}[\nabla^2u + (\nabla u)^2]. \eqno(1.19)$$
Therefore (2.13), (2.14) corresponds to
$$\tilde R_{ij} = +2 u_{;i}u_{;j}, \eqno(2.20)   $$
and
$$\tilde \nabla^2u = 0.\eqno(2.21)$$
Eq.(2.20) and Eq.(2.21) express the static field equations in the form
that is most convenient for further analysis.

The equations do not simplify so dramatically in the stationary case because
the Killing vector will not be hypersurface orthogonal. However, some
significant simplification can  still be achieved. We can no longer force the
conjugate momentum $(\pi^{ij})$ to vanish entirely but we can always choose our
slice so that $g_{ij}\pi^{ij} = 0 $, at least near infinity (Bartnik,
Chru\'sciel and \'O Murchadha 1990) (which is all we are interested in).

Taking the trace of Eq.(2.3) gives
$$ -N g^{-{1 \over 2}}g_{ab}\pi^{ab} + 2\nabla_aN^a = 0, \eqno(2.22)$$
which immediately gives
$$\nabla_aN^a = 0. \eqno(2.23) $$
The trace of Eq.(2.4) gives
$${1 \over 2}Ng^{{1 \over 2}}R - {1 \over 2}Ng^{-{1 \over 2}}\pi_{ab}\pi^{ab}
-2g^{{1 \over 2}}\nabla^2N - 2N_{a:b}\pi^{ab} = 0. \eqno(2.24)$$
Using the Hamiltonian constraint (2.1) we can eliminate the first two terms to
give
$$g^{{1 \over 2}}\nabla^2N = -N_{a:b}\pi^{ab} \eqno(2.25)$$
If we multiply (2.3) by $\pi^{ij}$ we get
$$Ng^{-{1 \over 2}}\pi_{ab}\pi^{ab} = -N_{a:b}\pi^{ab}.\eqno(2.26)$$
If we now substitute (2.26) into (2.25) to give
$$\nabla^2N = Ng^{-{1 \over 2}}\pi_{ab}\pi^{ab} = NR. \eqno(2.27)$$
This is the well-known equation for the choice of lapse which preserves the
maximal slicing $(g_{ab}\pi^{ab} = 0)$ condition. Finally, we can substitute
(2.27) into (2.4) to simplify it and give
$$Ng^{{1 \over 2}}R_{ab} = g^{{1 \over 2}}N_{;ab} -2Ng^{-{1 \over 2}}\pi_a^m
\pi_{mb} + (\pi_{ab}N^m)_{;m} - N_{a;m}\pi^m_b - N_{b;m}\pi_a^m.\eqno(2.28)$$
\vskip 1cm
\centerline{\bf III Weakly decaying static solutions}
\vskip 0.5cm
Following from Section II, we seek a metric g and a function N satisfying
$$NR_{ab} = N_{;ab} \eqno(3.1)$$
$$\nabla^2N =\ ^{(3)}R = 0 \eqno(3.2) $$
We now need to define suitable function spaces in which to define them. We
choose to use weighted Sobolev spaces. Let us define:
\proclaim Definition 1. $\sigma = (1 + |x^2|)^{{1 \over 2}}$.

\proclaim Definition 2. $H_{s, \delta}(^3R/B), s\in N, \delta \in \Re$, is
the class of all functions $f$ on $^3R/B$ possessing weak derivatives up to
order $s$ such that $\sigma^{-\delta -{3 \over 2} +|\alpha|}D^{\alpha}f$, for
each $|\alpha| \leq s$, is square integrable on $^3R/B$.

 This forms a Hilbert space with norm
$$\|f\|_{H_{s, \delta}} = \sum_{|\alpha| \leq s}\left\{ \int_{^3R/B}
\sigma^{-2\delta - 3 + 2|\alpha|}|D^{\alpha}f|^2 dv \right\}.$$

{\bf Note:} This definition of a weighted Sobolev space is the standard one
(as in Choquet-Bruhat and Christodoulou 1981) except that (following Bartnik
1986) we have subtracted 3/2 from the old definition of $\delta$ and changed
sign. Now a function in $H_{s, \delta}$ has (more or less) classical blowup
like $r^{+\delta}$.

One standard result (Choquet-Bruhat and Christodoulou 1981) we will frequently
use is the:
\proclaim  Multiplication Theorem. Pointwise multiplication is a
continuous bilinear map
$$(f_1, f_2) \rightarrow f_1f_2$$
$$H_{s_1, \delta_1} \times H_{s_2, \delta_2} \rightarrow H_{s, \delta}$$
if $s_1$, $s_2 \geq s$, $s < s_1 + s_2 - 3/2$, $\delta > \delta_1 + \delta_2$.

We wish to consider static solutions that are asymptotically flat in the sense
that the metric (in some coordinate frame) approaches the flat cartesian metric
at infinity. We also assume that the killing vector approaches the standard
time-translation killing vector in the same frame. In other words, we wish to
consider solutions to (3.1), (3.2), (3.3) on a manifold with topology $^3R/B$,
having the following properties:
$$   g - \delta \in H_{s, \delta}\;, \quad s \geq 3\;,\; \delta < 0
\eqno(3.4a)$$
 $$  N \rightarrow 1 \hbox{ at }\infty \;. \eqno(3.4b)$$
Condition (3.4a) on the metric says that it approaches the cartesian metric
$\delta$ at infinity, but we do not care how slowly; the condition (3.4b) on
$N$ (the lapse, the length of the Killing vector) says that it goes to a
time-translation at infinity. Standard theorems on the laplacian (see e.g. \'O
Murchadha (1986)) allow us to deduce immediately from (3.2) that $N$ must
satisfy $$N - 1 + {M \over r} \in H_{s+1, \delta-1} \;, \eqno(3.5)$$
where $M$ is some constant.

We know that $N - 1$ satisfies $\nabla^2(N - 1) = 0$ and when one explicitly
writes this out it becomes clear that the leading part of $N - 1$ must be a
harmonic function of the flat space laplacian. The residue then falls off
faster than 1/r. Note also that $N$ (again because it satisfies the laplacian)
is one degree more differentiable than g.

Now we know that $N$ is smooth and well-behaved [in particular we need that $N$
is bounded away from zero in a neighbourhood of infinity, we can use the
standard embedding theorem (Choquet-Bruhat and Christodoulou 1981) to prove
this], we can use it as a conformal factor, $\tilde g = N^2g = e^{2u}g$.
$\tilde g$ can be shown to be a riemannian metric which satisfies
$$\tilde g - \delta \in H_{s,\delta},\;\;\;\;s \geq 3,\; \delta < 0\;,
\eqno(3.6)$$
outside some ball $B'$.

The field equations take the form ((2.20), (2.21))
$$\tilde R_{ij} = +2 u_{;i}u_{;j}, \eqno(3.7)   $$
and
$$\tilde \nabla^2u = 0,\;\;\;\;u \rightarrow 0 \hbox{ at } \infty\;.
\eqno(3.8)$$
again, the standard theorem for the laplacian on weighted Sobolev spaces gives
$$u - {A \over r} \in H_{s+1,\delta -1}\;,\eqno(3.9)$$
for some constant $A$. (This also follows directly from (3.5) and $u = \ln N$.)

When this is substituted into (3.7) we see that the Ricci curvature of the
conformally transformed space falls off like 1/$r^4$. We actually get
$$\tilde R_{ij} - 2A^2{x^ix^j \over r^6} \in H_{s, \delta - 4},\eqno(3.10)$$
where $A$ is the same constant as in (3.9).
Since in three dimensions, Ricci is equivalent to Riemann, this means that the
Riemann curvature also falls off like 1/$r^4$. This, in turn means that the
space must be flat to order 1/$r^2$. When we return to the physical space by
multiplying by $N^{-2}$ we introduce a 1/r term into the metric but this term
is a pure conformal factor of the form 2M/r. Therefore the
physical three-space must be pure Schwarzschild to order 1/r, $g_{ab} = (1 +
2M/r)\delta_{ab} + {\cal O}(1/r^2)$.

This argument can be made more precise by introducing harmonic coordinates in
the conformal space. We seek three functions $\phi_1, \phi_2, \phi_3$
satisfying
$$\tilde \nabla^2\phi_1 = 0\;, \;\;\; \phi_1 \rightarrow x \hbox{ at }\infty
\;, \eqno(3.11)$$
and similarly for $\phi_2$ and $\phi_3$. Such functions exist and belong to
$H_{s+1,1}$. We use these as our new coordinates near infinity (the details can
be seen, for example, in \'O Murchadha 1986) and the transformed metric
$\tilde g'$  will belong to the same space as $\tilde g$, $\tilde g' - \delta
\in H_{s, \delta}$. Writing  $\tilde h_{ab} = \tilde g'_{ab} - \delta_{ab},$
we get $$\tilde R_{ab} = \tilde \nabla^2 \tilde h_{ab} + \Gamma\Gamma
.\eqno(3.12)$$ This is the standard simplification of the Ricci tensor that
occurs on using harmonic coordinates. $\Gamma$ is essentially just the first
derivative of the metric so it is easy to show $\Gamma \in H_{s-1, \delta
-1}$. The multiplication theorem can now be used to show $\Gamma\Gamma \in
H_{s-1, 2\delta -2}$. This is where we need to use $s \geq 3$, otherwise the
derivatives go bad.
We rewrire (3.12) as
$$\tilde \nabla^2 \tilde h_{ab} = \tilde R_{ab} - \Gamma\Gamma. \eqno(3.13)$$
Since we know, from (3.10), that the Ricci tensor is
both smoother and has faster falloff than $\Gamma\Gamma$, we see that (3.13)
forces $\tilde h_{ab}$ into $H_{s+1, 2\delta}$. Therefore we improve both
the differentiability and the falloff of the metric. We iterate this
argument $l$ times where $l$ satisfies $2^l\delta < -1$. At this point we
find that $\Gamma\Gamma$ (classically) falls off faster than 1/$r^3$. At
this stage when we invert the laplacian (see \'O Murchadha 1986) we pick up a
harmonic function of the flat-space laplacian. In other words we have
$$\tilde h_{ab} - {A_{ab} \over r} \in H_{ls,-\epsilon -1}, \eqno(3.14)$$
where $A_{ab}$ are six (apparently independent) constants. However, we also
require that the metric be harmonic, i.e., $2\tilde h_{ab,b} - \tilde
h_{bb,a} = 0$. When we apply these conditions to (3.14) we get a set of
relationships between the $A_{ab}$'s which cannot be satisfied. Therefore
they must all vanish and $\tilde h_{ab}$ must fall off faster than 1/r. We
can now continue the iteration and show that $\tilde h_{ab}$ must fall off
like 1/$r^2$. At this stage we have improved the falloff and
differentiability sufficiently so that we can appeal to Beig and Simon and get
that $\tilde h_{ab}$ must be analytic.

The content of this Section can therefore be summarized as
\proclaim Theorem.
Given an asymptotically flat Riemannian three-metric $g - \delta \in
H_{3,\delta}$, $\delta < 0$, which is the metric on a slice transverse to the
killing vector in a static space-time which is vacuum near infinity, and given
that the length of the killing vector goes to a constant at infinity, it
follows that the metric must be analytic  and Schwarzschildian at infinity.

{\bf Remarks:}(i) We need three weak derivatives because we have to estimate
terms like $(g_{ab,c})^2$ and have them smoother than $g_{ij,kl}$ so as to
improve the differentiability. If we were willing to consider non-integral
differentiability what we require is $x - 2 < 2(x - 1) - 3/2$. This implies $x
> 3/2$. In particular s = 2 will work.

\hskip 1.5cm  (ii) The $\delta < 0$ condition classically means that we are
willing to consider any metric which decays to flat space like
1/$r^{\epsilon}$ for any $\epsilon$. Obviously one would like to replace
this with `going flat' and not require any kind of power-law decay. It is
difficult to see how this might be achieved; none of the present battery of
weighted spaces (classical, H\"older \dots ) seem suitable.

\hskip 1.5cm (iii) The condition $N \rightarrow 1$ at infinity can be
significantly relaxed. All we need to do is to eliminate the harmonic
functions of the laplacian. Therefore all we need is that N grows slower
than $r$ at infinity. The question of the behaviour of the solution if one
places no restriction at all on N must be analyzed separately.
\vskip 1cm
\centerline{\bf IV: Weakly decaying stationary solutions}
\vskip 0.5cm
As we showed in Section II, the stationary equations are equivalent to the
existence of a Riemannian metric $g$, a tracefree symmetric tensor $P$ (we
define $P^{ab} = g^{-{1 \over 2}}\pi^{ab}$ to get rid of the determinants of
$g$), a scalar $N$ (the lapse, the length of the part of the killing vector
orthogonal to the slice) and a vector $N^i$ (the shift, the part of the
killing vector in the slice) satisfying
$$^{(3)}R = P^{ab}P_{ab}.\eqno(4.1)$$
$$\nabla_aP^{ab} = 0.\eqno(4.2)$$
$$\nabla^2N = NR. \eqno(4.3)$$
$$\nabla^2N = - N_{a;b}P^{ab}. \eqno(4.4)$$
$$2NP_{ab} + N_{a;b} + N_{b;a} = 0. \eqno(4.5)$$
$$g_{ab}P^{ab} = 0.\eqno(4.6)$$
$$\nabla_aN^a = 0. \eqno(4.7)$$
$$NR_{ab} = N_{;ab} - 2NP_a^mP_{bm} + P_{ab;m}N^m - N_{a;m}P^m_b -
N_{b;m}P^m_a .\eqno(4.8)$$
As in Section III we want to make some (weak) assumptions about the asymptotic
behaviour of the metric, the extrinsic curvature and of the killing vector. We
wish to assume
$$g_{ab} - \delta_{ab} \in H_{3,\delta}, \eqno(4.9)$$
$$N \rightarrow 1, \hskip 0.5cm N^a \rightarrow 0 \hbox{   at }\infty.
\eqno(4.10)$$
Equation (4.9) gives us that $^{(3)}R \in H_{1,\delta-2}$. This essentially
forces (from eqn.(4.1))
$$P^{ab} \in H_{2,{\delta \over 2}-1}.\eqno(4.11)$$
We assume that this holds. Equation (4.1) also tells us that $^{(3)}R \geq 0$
and this means that $N$ (from eqn.(4.3)) satisfies a maximum principle from
which we find that (Theorem 3.5, Christodoulou and \'O Murchadha 1981)
$$ N - 1 \in H_{3,\delta}.\eqno(4.12)$$
Taking the divergence of the tracefree part of (4.5) gives
$$-2N_{;a}P^a_b = [N_{a;b} + N_{b;a} - {2 \over 3}\nabla_cN^cg_{ab}]^{;a}.
\eqno(4.13)$$
The multiplication theorem, together with eqns.(4.11) and (4.12), gives us
that $2N_{;a}P^a_b \in H_{2,{3\delta \over 2}-2}$. The right-hand-side of
(4.13)
is the divergence of the conformal killing form of the vector $N^a$. This is
a very nice second order linear elliptic operator. We immediately get
(Lemma 3.1, Christodoulou and \'O Murchadha 1981)
$$N^a \in H_{3,{3\delta \over 2}}. \eqno(4.14)$$
We can substitute (4.11) and (4.14)  into (4.4) to get $\nabla^2N \in
H_{2,2\delta -2}$. In turn this gives
$$ N - 1 \in H_{4,2\delta}.\eqno(4.15)$$
Comparing with (4.12), we see that we have improved both the smoothness and
the falloff of $N$. When (4.15) is substituted into (4.13) we now get that
$2N_{;a}P^a_b \in H_{2,{5\delta \over 2}-2}$. In turn this gives
$$N^a \in H_{3,{5\delta \over 2}}. \eqno(4.16)$$
We can iterate this procedure and keep improving the falloff of both $N$ and
$N^a$. There exists an integer $n$ such that $n\delta < -1$. We get
$\nabla^2N \in H_{2,n\delta -2}$. Then we find
$$ N - 1 + C/r \in H_{4,n\delta},\eqno(4.17)$$
where $C$ is some constant. We also find that $N^a$ falls off like 1/$r$. Thus
we can show that the killing vector must be quite regular at infinity.

We now need to show that the metric and extrinsic curvature have similar
properties. The easiest way to proceed is to imitate Beig and Simon. We have
shown that $N$ and $N^a$ are well-behaved at infinity. Therefore we see that
$\lambda = N^2 - N^aN_a$ is also regular, so that
$$ \lambda - 1 + C/r \in H_{3,-1-\epsilon},\eqno(4.18)$$
where $\lambda$ is the length of the killing vector. Also regular are
$\sigma_a = N_a/\lambda$ and $\tilde{g}_{ab} = \lambda (g_{ab} +
\sigma_a\sigma_b)$ with
$$\sigma_a \in H_{3,-1}, \eqno(4.19)$$
$$\tilde{g}_{ab} - \delta_{ab} \in H_{3,\delta}. \eqno(4.20)$$
It can be shown that $\tilde{g}_{ab}$ is a regular riemannian metric, at least
in a neighbourhood of infinity.

The next step is to realise that $\omega_a$
$$\omega_i = -\lambda^2\epsilon_{ijk}\tilde{D}^j\sigma^k \eqno(4.21)$$
is curlfree, and can be replaced by a scalar $\omega$, $\omega_a =
\tilde{D}_a\omega$. Finally, Beig and Simon replace $(\lambda, \omega)$ by
$\phi_A$ ($A = 1, 2$)
$$\phi_1 = {1 \over 4}\lambda^{-1}(\lambda^2 + \omega^2 - 1), \hskip 1cm
\phi_2 = {1 \over 2}\lambda^{-1}\omega. \eqno(4.22)$$
The field equations now read
$$\tilde{\nabla}^2\phi_A = 2\tilde{R}\phi_A \eqno(4.23a)$$
$$\tilde{R}_{ab} = 2\tilde{D}_a\phi_A\tilde{D}_b\phi_A -
\tilde{D}_a\Sigma\tilde{D}_b\Sigma, \eqno(4.23b)$$
with $\Sigma = {1 \over 2}(1 + 4\phi_A\phi_A)^{1 \over 2}$.

Now we can iterate on (4.23) without difficulty and show that we can improve
both the decay rates and the differentiability of $(\tilde{g}_{ab}, \phi_A)$
essentially without limit. Eventually we reach high enough differentiability
so that we can replace the weak derivatives with ordinary derivatives; we then
can use Beig and Simon to prove analyticity.

Thus we have proven:
\proclaim Theorem.
Given a solution to the Einstein equations which is asymptotically flat,
vacuum outside a region of compact support and possesses a timelike
killing vector in the exterior region, such that $g_{\mu\nu} -
\eta_{\mu\nu} \in H_{4,\delta}(R^4 - B)$ for any $\delta < 0$ in the natural
coordinates   (where
$\partial/\partial t$ is the killing vector) the metric must be
analytic and Schwarzschildian at infinity.

 \vskip 1cm \centerline{\bf
References} \vskip 0.5cm \parindent=0pt *Current address: Theoretical
Astrophysics, California Institute of Technology, Pasadena, Calif.  U.S.A.\par
Arnowitt, R., Deser, S., Misner, C. W. 1962 The dynamics of general
relativity. In {\it Gravitation: An introduction to current research} (ed. E.
Witten), pp. New York: J. Wiley \& Sons.\par
Bartnik, R. 1986 The mass of an asymptotically flat manifold. {\it Comm. Pure
and Appl. Maths.} {\bf 39}, 661-693.\par
Bartnik, R., Chru\'sciel, P. T., \'O Murchadha, N. 1990 On maximal surfaces
in asymptotically flat space-times. {\it Commun. Math. Phys.} {\bf 130},
95-109.\par
Beig, R., Simon, W. 1980 The stationary gravitational field near spacelike
infinity {\it G.R.G.} {\bf 12}, 439-451  \par
Beig, R., Simon, W. 1981 On the
multipole expansion for stationary space-times {\it Proc. R. Soc. Lond.} {\bf
A 376}, 333-341.\par Choquet-Bruhat, Y., Christodoulou, D. 1981 Elliptic
systems in $H_{s,\delta}$ systems which are euclidean at infinity. {\it Acta
Math.} {\bf 146}, 129-150.\par
Christodoulou, D., \'O Murchadha, N. 1981 The boost problem in general
relativity. {\it Commun. Math. Phys.} {\bf 80}, 271-300.\par
Chru\'sciel, P. T. 1986a Boundary conditions at spatial infinity from a
Hamiltonian point of view. In {\it Topological properties and global
structure of space-time} (eds. P. G. Bergmann \& V. de Sabbata), pp. 49-59,
New York: Plenum Press.\par
Chru\'sciel, P. T. 1986b A remark on the positive-energy theorem. {\it Class.
Quantum Grav.} {\bf 3}, L115-L121.\par
\'O Murchadha, N. 1986  Total energy-momentum in general relativity
. {\it J. Math. Phys.} {\bf 27}, 2111-2128.

 \bye